\documentclass[prl,twocolumn,superscriptaddress,prl]{revtex4-1}

\usepackage{amsmath}
\usepackage{amssymb}
\usepackage{xspace}
\usepackage{graphicx}
\usepackage{grffile}
\usepackage{nicefrac}
\usepackage{color}

\graphicspath{{figs/}}

\begin{document}

\title{Electron dichotomy on the SrTiO$_3$ defect surface augmented by many-body effects}

\author{Frank Lechermann}
\affiliation{I. Institut f{\"u}r Theoretische Physik, Universit{\"a}t Hamburg, 
D-20355 Hamburg, Germany}
\affiliation{Institut f\"ur Keramische Hochleistungswerkstoffe, Technische
Universit\"at Hamburg-Harburg, D-21073 Hamburg, Germany}
\author{Harald O. Jeschke}
\affiliation{Institut f{\"u}r Theoretische Physik,
Goethe-Universit\"at Frankfurt, Max-von-Laue-Str. 1, 60438 Frankfurt am Main, Germany}
\author{Aaram J. Kim}
\affiliation{Institut f{\"u}r Theoretische Physik,
Goethe-Universit\"at Frankfurt, Max-von-Laue-Str. 1, 60438 Frankfurt am Main, Germany}
\author{Steffen Backes}
\affiliation{Institut f{\"u}r Theoretische Physik,
Goethe-Universit\"at Frankfurt, Max-von-Laue-Str. 1, 60438 Frankfurt am Main, Germany}
\author{Roser Valent{\'i}}
\affiliation{Institut f{\"u}r Theoretische Physik,
Goethe-Universit\"at Frankfurt, Max-von-Laue-Str. 1, 60438 Frankfurt am Main, Germany}


\begin{abstract}
In a common paradigm, the electronic structure of condensed matter is divided into 
weakly and strongly correlated compounds. While conventional band theory usually works 
well for the former class, many-body effects are essential for the latter. 
Materials like the familiar SrTiO$_3$ compound that bridge or even abandon this 
characterization scheme are highly interesting. Here it is shown by means of combining 
density functional theory with dynamical-mean field theory that oxygen vacancies on the 
STO (001) surface give rise to a dichotomy of weakly-correlated $t_{2g}$ low-energy 
quasiparticles and localized 'in-gap' states of dominant $e_g$ character with subtle 
correlation signature. We furthermore touch base with recent experimental work and 
study the surface instability towards magnetic order.
\end{abstract}

\maketitle
The SrTiO$_3$ (STO) compound has long been known for its paraelectric~\cite{coc60}, 
semiconducting~\cite{fre64} and superconducting properties~\cite{scho64}. Renewed 
interest is fostered by the findings of a conducting two-dimensional electron system (2DES)
on the STO surface~\cite{san11,mee11}, of intricate magnetic response~\cite{ric14}, as well 
as due to its intriguing role as a prominent building block in novel oxide 
heterostructures (\cite{oht02,oka04}, see e.g.~\cite{zub11,hwa12} reviews). In this respect, 
several experiments~\cite{tan93,aiu02,san11,mee11,mck14} point towards the importance of 
oxygen vacancies for the plethora of physics emerging from the inconspicuous bulk band 
insulator. 

Stoichiometric strontium titanate is rather unsusceptible to electronic 
correlations due to nominal Ti$^{4+}$$(3d^0)$ valence. However, doping transforms STO 
into a material with potential for salient signatures of correlation effects. Usually 
the competition between electron localization and itinerancy in materials can be traced back 
to the interacting many-body system {\sl at stoichiometry}. In doped STO, in contrast, defects 
have to build up the general correlated electronic structure from localized states  
affecting the low-energy quasiparticle (QP) nature beyond a conventional Anderson-model 
perspective. 

The orbital character, filling, correlation strength and mobility of the key electronic 
states in doped STO is of main interest. In principle, electron doping as introduced by 
oxygen vacancies could fill the empty Ti-$3d(t_{2g})$ states at low energy. However creating 
an oxygen vacancy (OV) breaks the bond between Ti-$3d(e_g)$ and O-$2p$. It has been 
shown~\cite{luo04,hou10,mit12,pav12,lin13,lec14,jes15,alt15} that as a result also $3d(e_g)$ 
spectral weight appears just below the Fermi level $\varepsilon_{\rm F}$ of the established 
metallic state. Investigations based on density functional theory (DFT) identify this $e_g$ 
weight as associated with an 'in-gap' level due to the vacancy-induced crystal-field 
lowering. Depending on the local structural deformation, Ti charging, vacancy 
concentration and energy distance to $\varepsilon_{\rm F}$, that spectral feature may 
also be interpreted as a small polaron. 
The latter is here defined by the localization of one electron at an isolated Ti site, i.e. 
forming Ti$^{3+}$, with sizable distortion of the surrounding oxygen octahedron and
binding energy $\sim -1$eV~\cite{die03}. Recent DFT+ Hubbard $U$ calculations for the 
TiO$_2$ rutile and anatase surface reveal that local Coulomb interactions are relevant for 
small-polaron formation~\cite{set14}.

The STO surface is readily made accessible to the creation of OVs, e.g. via in-situ crystal 
fracture~\cite{san11} or by exposure to UV light~\cite{mee11}. Furthermore (angle-resolved) 
photoemission spectroscopy (ARPES) is ideally suited to probe the surface electronic 
structure. It reveals $t_{2g}$-like low-energy bands and a prominent high-energy peak at 
$-1.3$eV~\cite{aiu02,mee11,mck14}. While the $t_{2g}$-band dispersion is well described 
in DFT-based methods~\cite{she12,jes15,alt15}, the characterization of the high-energy feature is 
more intriguing. Depending on the computational setup, the in-gap $e_g$ weight from 
DFT+U~\cite{jes15,alt15} matches with the corresponding experimental peak position 
below $\varepsilon_{\rm F}$. Though Kohn-Sham theory with static Hubbard-$U$ correlations
provides relevant insight~\cite{jes15,alt15}, so far a clear-cut many-body study of the 
interplay between high- and low-energy electron states on the oxygen-deficient STO 
surface has been missing. Also in a small-polaron picture the (orbital-dependent) role of 
electronic correlations, especially beyond static modelings, remains open.

In this work, we employ a combination of DFT with dynamical mean-field theory (DMFT) to
investigate the role of correlations due to isolated oxygen-vacancy defects on the STO (001)
surface. We find that many-body effects within charge self-consistent DFT+DMFT reveal novel
features of the high-energy $-1.3$eV peak beyond the means of static correlation schemes.
Localized $e_g$-like states with minor $t_{2g}$ intermixing and itinerant states of exclusive 
$t_{2g}$ nature with strong $xy$ polarization coexist. An interplay of renormalized crystal fields 
and finite-frequency parts of the electronic self-energies drives this electronic dichotomy.

\paragraph{Surface defect structure.}
\begin{figure}[t]
\begin{center}
(a)\hspace*{0.02cm}\includegraphics*[width=3.75cm]{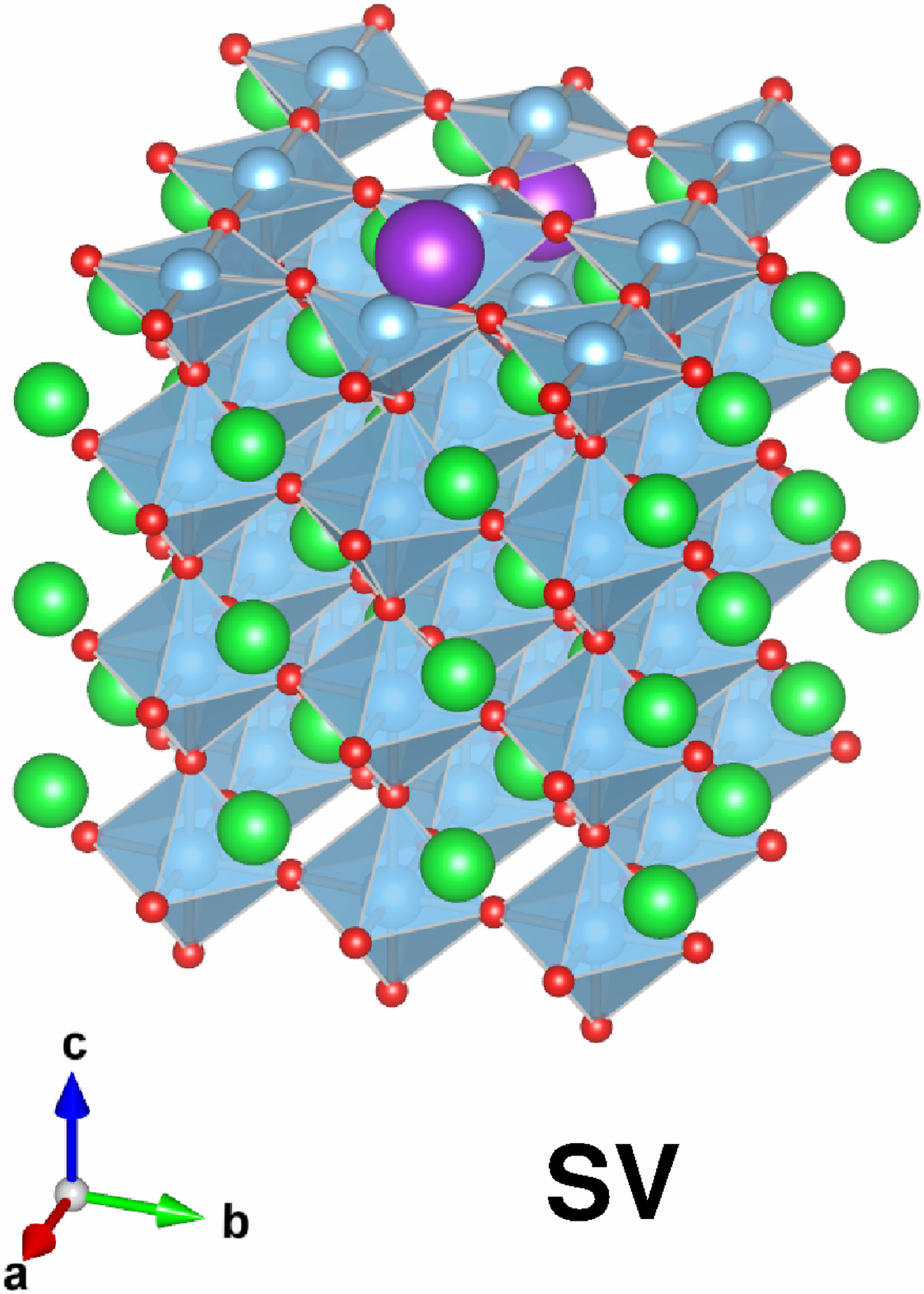}\hspace*{-0.2cm}
(d)\hspace*{0.02cm}\includegraphics*[width=3.75cm]{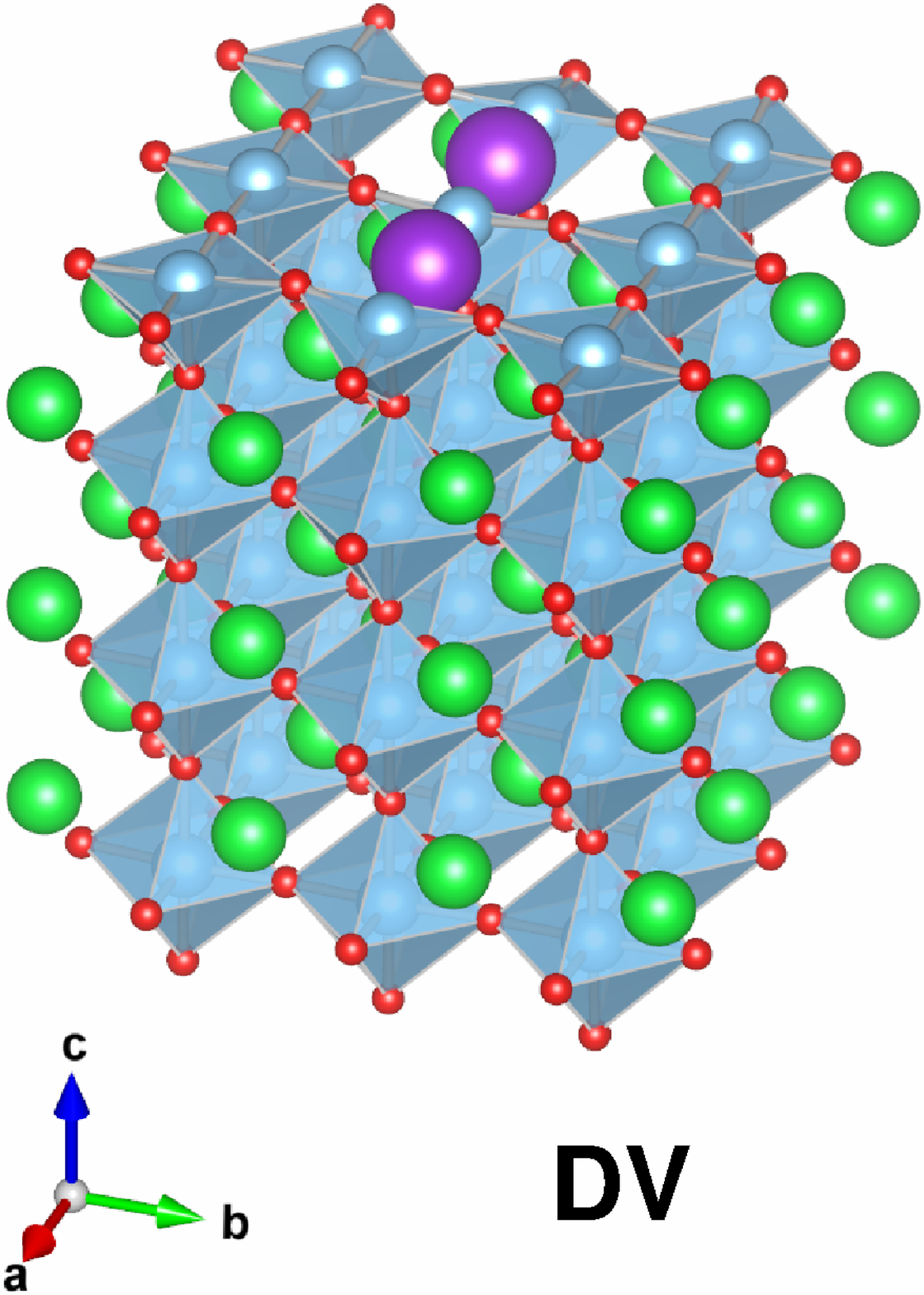}\\[0.15cm]
\includegraphics*[height=4.8cm]{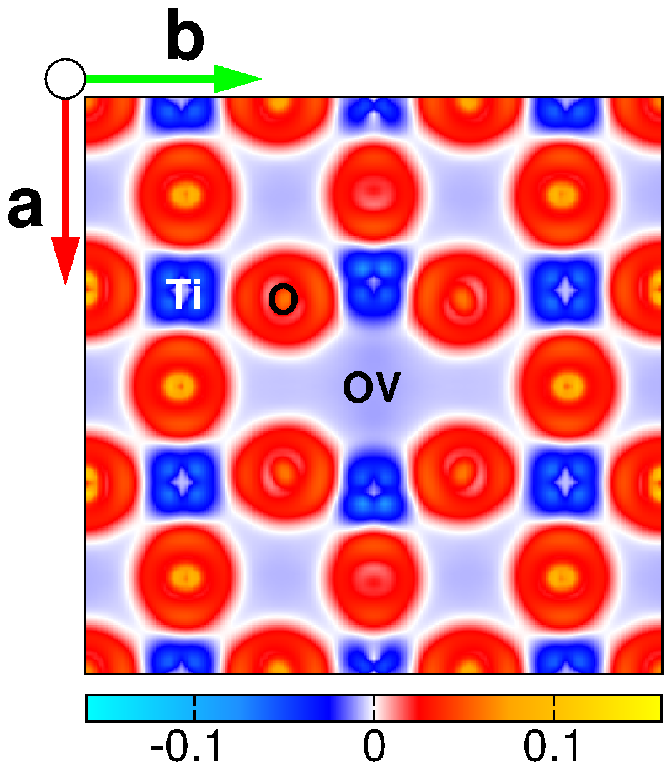}\hspace*{0.1cm}
\includegraphics*[height=4.8cm]{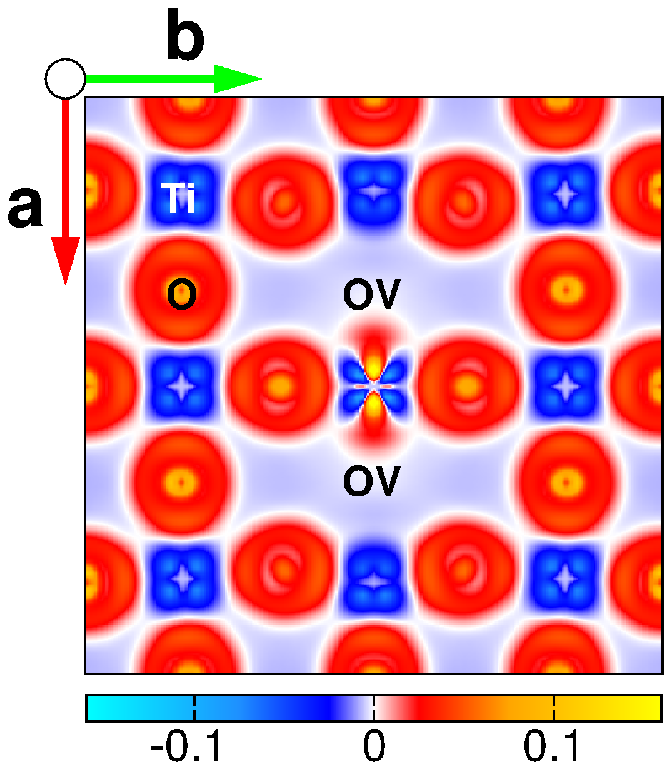}\\
\hspace*{-3cm}(b)\hspace*{4cm}(e)\\[0.15cm]
\includegraphics*[height=4.8cm]{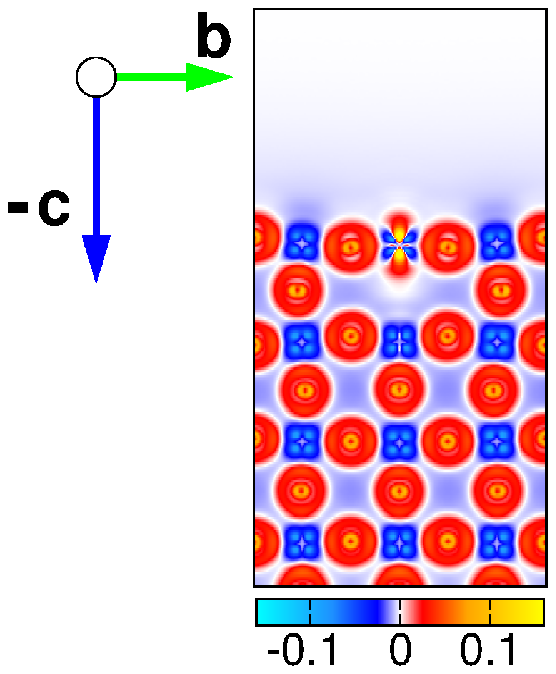}\hspace{0.2cm}
\includegraphics*[height=4.8cm]{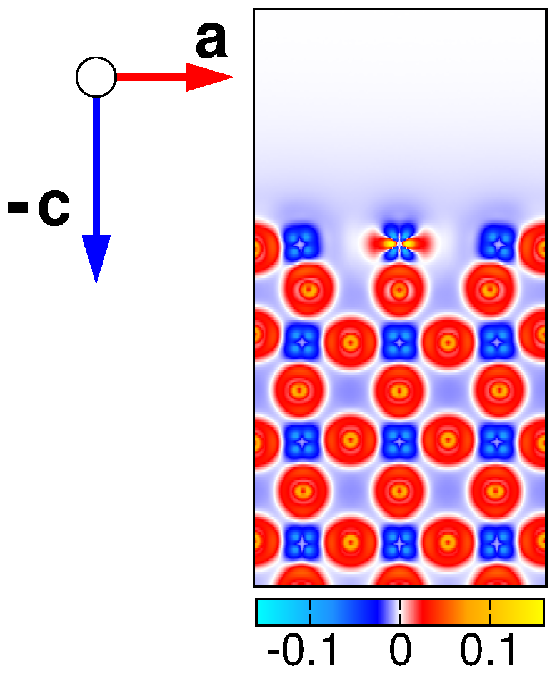}\\
\hspace*{-3cm}(c)\hspace*{4cm}(f)
\end{center}
\caption{(color online) 
DFT+DMFT bonding charge density $\rho_{\rm b}$$\equiv$$\rho_{\rm slab}$$-$$\rho_{\rm atomic}$
for the SrTiO$_3$ oxygen-defect surface in the (a-c) two-single-vacancy structure (SV) and 
(d-f) double-vacancy structure (DV). Sr (green), Ti (lightblue), O (small red) and OV 
(large violet). (b,e) Top view on surface and (c,f) side view, cutting OV, respectively. 
The nominal Ti$^{4+}$(O$^{2-}$) ion looses(gains) charge compared to the free atom and 
thus appears blue(red).}\label{fig1}
\end{figure}
Periodic 3$\times$3$\times$4 slabs model the oxygen-deficient STO (001) 
surface~\cite{jes15}. Two defect architectures with a total of 180 atomic sites 
are investigated (see Fig.~\ref{fig1}a,d). In the first single-vacancies (SV) structure there 
is one OV in the surface layer and the second in the SrO layer below. The second structure bears 
a double-vacancy (DV) defect in the surface layer, i.e., two OVs in nearest-neighbor (NN) 
distance (for more details see the supplementary material).

\paragraph{DFT+DMFT approach.}
A correlated subspace of three effective $3d$ orbitals on each Ti site is 
obtained from the projected-local-orbital formalism~\cite{ama08,ani05,aic09,hau10,kar11}
(for details see the supplementary material). The Kohn-Sham problem within charge self-consistent 
DFT+DMFT~\cite{sav01,min05,pou07,gri12} is addressed by a mixed-basis pseudopotential 
code~\cite{mbpp_code}, employing the generalized-gradient approximation (GGA).
\begin{table}[b]
\begin{ruledtabular}
\begin{tabular}{c|rr|rrr}
     & $\varphi_{\rm SV}^{(1)}$ & $\varphi_{\rm SV}^{(2)}$ & $\varphi_{\rm DV}^{(1)}$ & 
$\varphi_{\rm DV}^{(2,2')}$ & $\varphi_{\rm DV}^{(3)}$   \\ \hline
GGA(PBE)   & 1.2 & 0.3 & 1.5 & 0.2 & $<$0.1  \\
DFT+DMFT   & 1.6 & 0.2 & 1.5 & 0.1 & 0.4\\
\end{tabular}
\end{ruledtabular}
\begin{eqnarray*}
\varphi_{\rm SV}^{(1,2)}&\sim&|z^2\rangle
\quad\mbox{above (1) and below (2) subsurface OV,}\nonumber\\
\varphi_{\rm DV}^{(1)}&\sim&-0.33|z^2\rangle+0.94|x^2-y^2\rangle
\quad\mbox{on embedded site,}\nonumber\\
\varphi_{\rm DV}^{(2,2')}&\sim&-0.39|z^2\rangle\pm 0.04|xz\rangle+0.92|x^2-y^2\rangle
\quad\mbox{next to OVs,}\nonumber\\
\varphi_{\rm DV}^{(3)}&\sim&|yz\rangle
\quad\mbox{on embedded site,}\nonumber\\
\end{eqnarray*}
\vspace*{-1.2cm}
\caption{Wannier-like orbital states with dominant filling, centered on Ti sites. Those 
states form the respective in-gap states. For degenerate $\varphi_{\rm DV}^{(2,2')}$ the 
electron-count sum is given.
\label{tab:levfill}}
\end{table}
Continuous-time quantum Monte Carlo~\cite{wer06,triqs_code} is used for the coupled 
single-site DMFT impurity problems. Local Coulomb interactions in 
Slater-Kanamori form are parametrized by a Hubbard $U$=3.5eV and a Hund's coupling 
$J_{\rm H}$=0.5eV. The correlated electron problem is solved at $\beta$=40eV$^{-1}$.
\begin{figure*}[t]
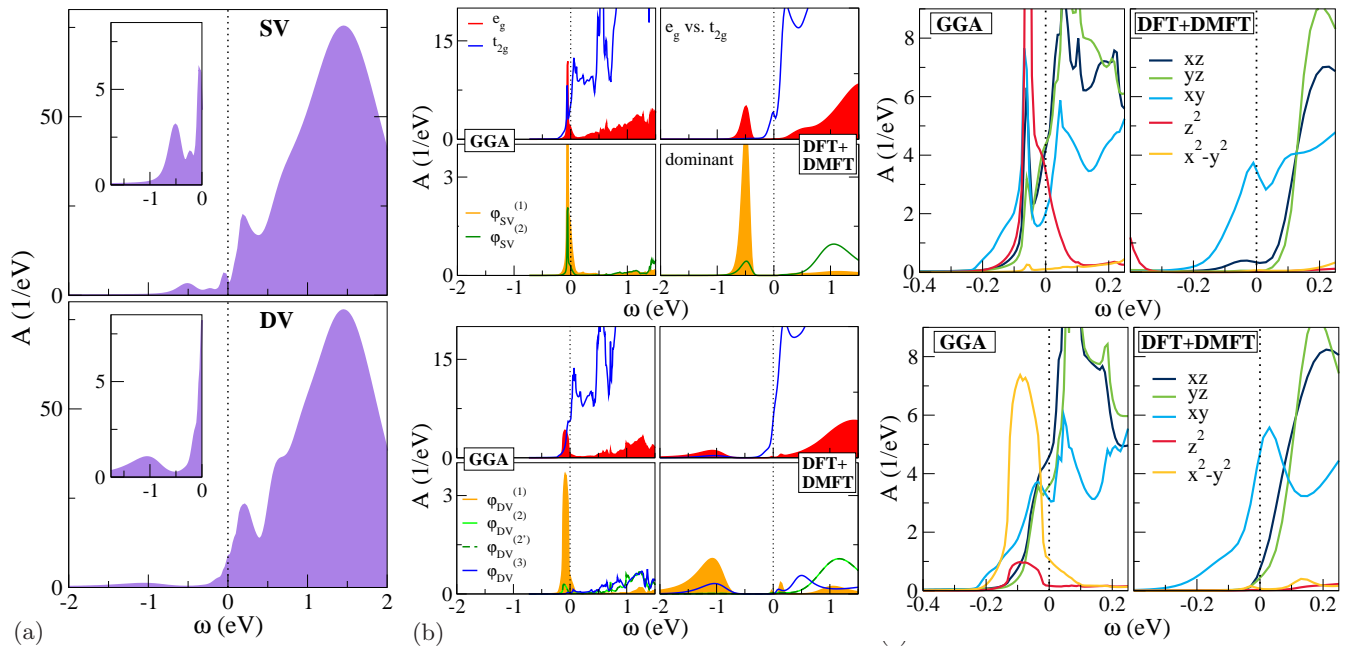

\hspace*{-0.2cm}
\parbox[c]{5.05cm}{(a)\hspace*{-0.45cm}\includegraphics*[height=8.4cm]{total-pm.eps}}
\parbox[c]{6.25cm}{\includegraphics*[height=4.2cm]{local-dd.eps}\\
(b)\hspace*{-0.45cm}\includegraphics*[height=4.2cm]{local-nn.eps}}
\parbox[c]{6cm}{\hspace*{-0.05cm}\includegraphics*[height=4.2cm]{low-dd.eps}\\
(c)\hspace*{-0.4cm}\includegraphics*[height=4.2cm]{low-nn.eps}}
\caption{(color online)
Paramagnetic spectral data for SV (top row) and DV (bottom row) defect cases. 
(a) Total spectral function from the DFT+DMFT 
Bloch Green's function of the conduction states.
(b) comparison with GGA concerning the $e_g$ vs. $t_{2g}$ occupation (top)
and the dominantly occupied Ti states (bottom).
(c) Low-energy Ti($3d$) weight compared to GGA.
}\label{fig2}
\end{figure*}

\paragraph{Paramagnetic electronic structure.}
Two OVs amount to a vacancy concentration of $c=0.02$ and electron-dope the slab
structures with four electrons, respectively. As a general finding in line with previous work, 
OV-induced localized defect states are indeed of dominant $e_g$ character. But still, both 
defect structures differ in their electronic characteristics, as visualized by the bonding 
charge density $\rho_{\rm b}$ in Fig.~\ref{fig1}. 
In the SV structure, only the surface Ti site above the OV in the SrO layer below harbors 
an effective $e_g$ level at lower energy. The surface-layer vacancy does not
allow for such states on nearby Ti sites, but the $t_{2g}$ levels are somewhat modified. 
On the other hand for the DV case, the double-OV defect only 
leads to effective $e_g$ on the Ti site between the two vacancies. Hence if double-vacancy
environments are available, low-energy $e_g$-like states are created on the embedded Ti site,
also at the cost of abandoning such defect levels at Ti sites near single vacancies. Note that
the subsurface oxygen vacancy in the SV case also creates such a double-vacancy scenario in
an effective way, since the vacuum mimics the role of the 'second vacancy'. 
Table~\ref{tab:levfill} provides the nature and the fillings of the dominantly occupied 
orbitals, i.e. those forming the in-gap states. Be aware that these Ti-based $\varphi$ orbitals are 
not of atomic but Wannier-like kind and their filling therefore includes charge from the surrounding
region. The GGA crystal-field splitting between the $e_g$-like levels and 
the $t_{2g}$ states is on average on the order of 1.3 eV. Notably in DFT+DMFT, in addition 
to the OV-introduced $e_g$-like fillings, the $t_{2g}(yz)$ orbital on the embedded surface Ti 
ion in the DV structural case becomes localized and significantly occupied. This
may be inferred from the confinement via OVs in $x$-direction, i.e. along the $a$-axis. Thus 
the true double-vacancy defect is effective in localizing additionally a $t_{2g}$ state through 
correlations. The quasi-double-defect involving the vacuum in the SV structure is not capable 
thereof. 

Both surfaces are metallic in GGA and DFT+DMFT. But the $k$-integrated spectral function with 
electronic correlations displays significant spectral weight transfer/shift compared to GGA 
(see Fig.~\ref{fig2}b,c). The total spectral functions with correlations in Fig.~\ref{fig2}a
display for the DV case a broadened high-energy peak at $-1.1$eV, i.e. close to experimental 
findings~\cite{tan93,aiu02,mee11,mck14}. A sharper peak at $\sim -0.5$eV occurs for the SV 
structure. As shown in the local analysis of Fig.~\ref{fig2}b, these in-gap spectral weights 
are mainly associated with $e_g$-like states already appearing on the GGA level. 
Electronic correlations are effective in broadening and shifting those towards higher energies. 
The real part of the self-energy $\Sigma(\omega+i0^+)$ renormalizes the GGA crystal field for 
$\omega=0$, while the imaginary part introduces finite-lifetime effects. Note that the DV 
high-energy satellite has in addition sizable $yz$ character from the embedded Ti site, 
whereas there is no substantial $yz$-orbital contribution at the associated GGA energy. 

In general, standard lower-Hubbard-band interpretations are not that readily applicable. 
The $e_g$-$t_{2g}$ hybridization is weak and already small interactions shift the $e_g$ 
spectral weight away from the Fermi level in the present isolated-defect cases. Furthermore, 
the $e_g$-like filling remains with and without correlations above a half-filled scenario
(cf. Tab~\ref{tab:levfill}). Both facts e.g. differ from a dense-defect limit of OVs in the 
LaAlO$_3$/SrTiO$_3$ interface, recently studied in DFT+DMFT~\cite{lec14}. There, a lower 
$e_g$-like Hubbard band from Ti$^{3+}$ was identified. Here the Wannier-like $3d$ occupation 
centered on OV-embedded Ti in the DV structure is close to integer filling of 
{\sl two} electrons, i.e. a local positive charging somewhat below Ti$^{3+}$ is likely. Note that 
Hubbard-like signatures may indeed be attributed to the $yz$ spectral contribution.

At the Fermi level, correlations are effective in suppressing the 
$e_g$ character from GGA (cf. Fig~\ref{fig2}b,c). Thus a clear dichotomy of 
only-localized and only-itinerant electrons sets in. Interestingly, 
for both structural cases the surface TiO$_2$ layer appears least conductive in DFT+DMFT. 
For the single-vacancy case the second transition-metal oxide layer contributes most to 
transport, while in the DV structure an equal share between the subsurface layers takes 
place. Moreover the $t_{2g}$-like QPs close to the Fermi level become strongly 
$xy$ polarized with correlations beyond GGA. Note that due to the small doping, strong
QP renormalization remains absent. The width $\Delta_{t_{2g}}$$\sim$0.25eV of this 
low-energy manifold agrees well with recent ARPES data~\cite{kin14,san14}.

\paragraph{Ferromagnetic electronic structure.}
Spin-splitted $t_{2g}$-derived low-energy bands have been observed in the 
2DES on the STO surface~\cite{kin14,san14}. Additionally, optically induced magnetism has
been directly reported for oxygen-deficient STO~\cite{ric14}. In a previous DFT+U
work~\cite{alt15} we showed that magnetism and Rashba effects compete at the doped-STO
surface and that the in-gap $e_g$ states acquire large magnetic moments. Here we study
correlation effects in the ferromagnetic (FM) state beyond DFT+U~\footnote{Our $T$=290K in the 
DMFT part should be well above a reasonable magnetic-ordering temperature for rather
isolated-defect induced magnetism. Due to the induced exchange splitting in the KS part,
information about details of low-temperature spin ordering can still be revealed in
charge self-consistent DFT+DMFT.}.
\begin{figure}[t]
\includegraphics*[height=4.25cm]{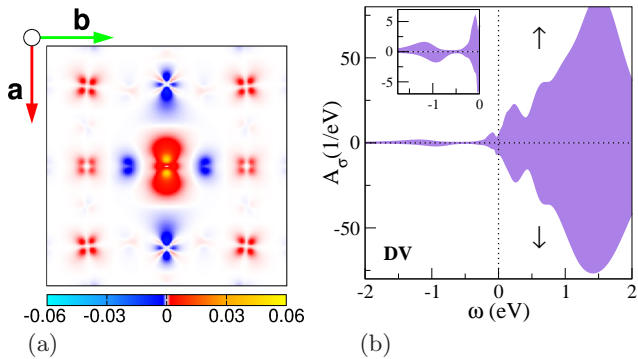}\hspace*{0.1cm}
\includegraphics*[height=4.25cm]{total-fm.eps}\\
\hspace*{-3cm}(a)\hspace*{4cm}(b)
\caption{(color online)
Ferromagnetic DV electronic structure in DFT+DMFT.
(a) Spin-resolved charge density in the $ab$ surface plane.
(b) Total Ti-based spectral function.}\label{fig4:totfmspec}
\end{figure}
When we allow for net spin polarization, the double-vacancy defect structure reveals a  
FM moment. No magnetic order is stabilized in the SV case.  
The DV magnetism is of strong local kind and spin splitting occurs mainly at the embedded 
Ti site (see Fig.~\ref{fig4:totfmspec}a). The ordered local moment amounts to 0.22$\mu_{\rm B}$, 
whereby the $\varphi^{(1)}$($\varphi^{(3)}$) orbital carries 0.14(0.08)$\mu_{\rm B}$. In our 
previous static DFT+U study for the same structure~\cite{alt15}, larger local moments up to 
0.9$\mu_{\rm B}$ were revealed. Thus quantum fluctuations and finite temperature are effective 
in reducing the ordered moment. Figure~\ref{fig4:totfmspec}a shows that in the DV surface layer, 
the NN oxygens and the OV-adjacent Ti sites counteract the embedded moment with minor 
antiferromagnetic alignment. 

The total spectral information in Fig.~\ref{fig4:totfmspec}b displays strong spin
splitting of about 0.25eV of the satellite peak at $-1.1$eV. The small $yz$ 
localized-orbital contribution from the embedded Ti site is nearly exclusively spin-up 
polarized. In contrast to some experimental data~\cite{san14} and our recent DFT+U 
work~\cite{alt15}, no strong spin splitting is detected in the low-energy $t_{2g}$ regime. 
This may be due to the elevated temperature, specific defect structure, or the present 
choice of the correlated subspace. Further work is needed to investigate this subtle issue 
with the ambitious DFT+DMFT framework in more detail.

\paragraph{Summary and discussion.}
Many-body and crystal-field effects induced by oxygen vacancies, here
described beyond static-correlation approaches within charge self-consistent DFT+DMFT, are
responsible for an apparent dichotomy in the electronic structure on the STO defect 
surface. Double-vacancy defects in the surface TiO$_2$ layer give rise to surface
localized $e_g$ dominated states as well as subsurface itinerant $t_{2g}$ metallicity.
The experimental $-1.3$eV high-energy satellite is identified as an $e_g$-derived in-gap level
that becomes broadened, shifted and dressed with minor Hubbard-band-like $t_{2g}(yz)$ weight by
a frequency-dependent self-energy. Note that this localized-$yz$ contribution may be unique to the 
specific architecture of the chosen DV structure. The signature of correlation is weaker in the 
single-vacancies defect structure, where the weakly-renormalized crystal-field character rules
the nature of the in-gap state. A direct connection to oxygen vacancies in bulk 
STO seems difficult at first sight, since there the surface symmetry breaking is absent. 
But based on earlier DFT studies~\cite{luo04,hou10} also revealing localized $e_g$-like states in 
bulk STO with such defects, the essential qualitative physics discussed here is believed to extend to 
the bulk regime. Moreover, in the recent DFT+DMFT study for LaAlO$_3$/SrTiO$_3$~\cite{lec14} a 
similar dichotomy has been revealed.

A comment on the comparison between the DFT+U and the DFT+DMFT approach to the problem is in order
The former method treating only static correlations already provides important insight~\cite{jes15}, 
also by including the Rashba physics~\cite{alt15}. 
However DFT+U mainly captures the renormalized crystal-field mechanism, fully established in that 
method only with symmetry breaking towards long-range magnetic order. This yields for doped STO a 
reasonable account of especially the integrated density of states. We here advance thereon by 
describing a true paramagnetic phase at finite temperature, where a competition between 
renormalized crystal fields and Hubbard-band-formation takes place. 

The small-polaron point of view is not studied in detail, albeit structural distortions 
around the OV defects gives room for such an interpretation, especially in the SV 
case. But metallic conductivity based on low-energy quasiparticles
sets in already for the present OV concentrations. Moreover, the possible deviation
from a clear Ti$^{3+}$ charge state, especially in the DV case, renders a direct 
small-polaron picture doubtful. Yet it cannot be excluded that polaronic transport 
is active in much more diluted scenarios.
On the other hand, it is generally expected that a further {\sl increase} of the OV 
concentration strengthens the Hubbard-like character of the high-energy satellite. The 
correlation-induced confining, here generated via the double-vacancy defect, will then be 
realized more effectively~\cite{lec14,jes15}. For larger numbers of OVs 
the $e_g$-$t_{2g}$ hybridization and -exchange will be such that $e_g$-like states 
participate in the low-energy transport~\cite{lec14,beh15}. Further study of that crossover and
inclusion of other possibly relevant ingredients such as the interplay of correlations 
and spin-orbit coupling beyond DFT+U or of improved non-local exchange-correlation 
treatments~\cite{roe14} are tasks for future theoretical work.

\begin{acknowledgments}
We thank G. Kresse for helpful discussions.
This research was supported by the Deutsche Forschungsgemeinschaft through FOR1346. 
Computations were performed at the University of Hamburg and the JUROPA/JURECA
Cluster of the J\"ulich Supercomputing Centre (JSC) under project number hhh08.

\end{acknowledgments}

\bibliography{bibextra}

\newpage

\begin{center}
{\LARGE Supplemental Information}
\end{center}
\section{Details on the Surface Structures}
For all calculations we utilize the STO experimental lattice constant $a$=3.905\AA. 
Periodic 3$\times$3$\times$4 slabs, a vacuum-region width of $5a$ and TiO$_2$ 
termination are used to model the oxygen-deficient STO (001) surface. 
Two defect architectures with two vacant oxygen sites and 178 remaining atomic sites, 
respectively, are investigated. The SV structure has one oxygen vacancy in the 
surface layer and the second in the SrO layer below, with separation distance 
$\sqrt{17}/2$\,$a$. The DV structure has a double-vacancy defect in the surface 
layer, i.e., two OVs in nearest-neighbor (NN) distance $a$. The respective 36 Ti 
sites in each structural case split into 16 symmetry-inequivalent shells. Both 
defect-surface slabes are structurally relaxed within DFT+U~[1]. While 
surface-layer OVs slightly repell the NN Ti ions, the subsurface vacancy 
attracts the Ti sites above and below.

\section{Details on the DFT+DMFT approach} 
The correlated subspace within the present work is build up by three effective $3d$ orbitals 
on each Ti site. In order to construct this subspace in a well-defined manner
we start from an initial full-$3d$-shell five-orbital projection using $5\times36=180$ Kohn-Sham (KS)
conduction states. After local orthogonalization the three most relevant projection 
functions are retained. Those are defined by having sizable occupation and/or weight
just above $\varepsilon_{\rm F}$, which proves reliable to reduce the number of important 
local Ti projections. Due to the symmetry-lowering defect structure, the effective functions 
are written as linear combinations of the original cubic ($e_g$, $t_{2g}$) orbitals. Finally 
the three-fold projection functions on each of the 36 Ti sites yield the correlated subspace 
by acting on $3\times36=108$ conduction KS states. It is important to realize that the resulting 
real-space orbitals of the correlated subspace are not highly localized at the Ti sites, such 
as true atomic orbitals or the derived projection functions. This due to the minimal projection of
108 localized functions onto 108 KS states, which here also excludes explicit O-based projection
functions. Hence the DMFT self-energies are associated with Wannier-like functions of some
extent. An interpretation of physical observables, such as the local electron filling, in
terms of atomic notions has therefore to be performed with care.

In the DFT part of the full correlated problem we use the generalized-gradient approximation 
(GGA) in the Perdew-Burke-Ernzerhof (PBE) form [2]. Due to symmetry, the DMFT part
has to account for 16 inequivalent single-site impurity problems. In the interfacing of DFT with
DMFT a double-counting correction is employed. We here put the fully-localized version~[3]
into practise. The complete correlated electronic structure is represented on a
4$\times$4$\times$1 $k$-point mesh. The maximum-entropy method serves for the 
analytical continuation from Matsubara space to retrieve the spectral data.\\

\noindent
$[1]$ H. O. Jeschke, J. Shen, and R. Valen{\'i}, New. J. Phys. {\bf 17}, 023034 (2015).\\
$[2]$ J. P.~Perdew, K.~Burke, and M.~Ernzerhof, Phys. Rev. Lett. {\bf 77}, 3865 (1996).\\
$[3]$ V.~I.~Anisimov, I.~V.~Solovyev, M.~A.~Korotin, M.~T.~Czy$\dot{\text{z}}$yk, and G.~A.~Sawatzky,
Phys. Rev. B. {\bf 48}, 16929 (1993).

\end{document}